\documentclass{raa}

\usepackage{longtable}
\usepackage{natbib}
\usepackage{amssymb}
\usepackage{graphicx,times}             
\input{epsf.sty}                        
\input{psfig.sty}                       

\begin{document}

   \title{A Possible Substellar Companion to the Intermediate-mass Giant HD 175679
}

   \volnopage{Vol.0 (200x) No.0, 000--000}      
   \setcounter{page}{1}          

   \author{Liang Wang
      \inst{1,2}
   \and Bun'ei Sato
      \inst{3}
   \and Gang Zhao
      \inst{1}
   \and Yujuan Liu
      \inst{1}
   \and Kunio Noguchi
      \inst{4}
   \and Hiroyasu Ando
      \inst{4}
   \and Hideyuki Izumiura
      \inst{5}
   \and Eiji Kambe
      \inst{5}
   \and Masashi Omiya
      \inst{3}
   \and Hiroki Harakawa
      \inst{3}
   \and Fan Liu
      \inst{1,2}
   \and Xiaoshu Wu
      \inst{1,2}
   \and Yoichi Takeda
      \inst{4}
   \and Michitoshi Yoshida
      \inst{6}
   \and Eiichiro Kokubo
      \inst{4}
   }

   \institute{
Key Laboratory of Optical Astronomy,
National Astronomical Observatories,
Chinese Academy of Sciences,
A20, Datun Road, Chaoyang District, Beijing 100012, China;
{\it wangliang@nao.cas.cn}\\
        \and
Graduate University of the Chinese Academy of Sciences,
19A Yuquan Road, Shijingshan District,
Beijing 100049, China\\
        \and
Department of Earth and Planetary Sciences,
Graduate School of Science and Engineering,
Tokyo Institute of Technology,
2-12-1 Ookayama, Meguro-ku,
Tokyo 152-8551, Japan\\
       \and
National Astronomical Observatory of Japan,
National Institutes of Natural Sciences, Mitaka,
Tokyo 181-8588, Japan\\
       \and
Okayama Astrophysical Observatory,
National Astronomical Observatory of Japan,
National Institutes of Natural Sciences, Asakuchi,
Okayama 719-0232, Japan\\
       \and
Hiroshima Astrophysical Science Center,
Hiroshima University, 1-3-1 Kagamiyama,
Higashi-Hiroshima 739-8526, Japan\\
}

   \date{Received~~2011 month day; accepted~~2011~~month day}

\abstract{
We report the discovery of a substellar companion around the intermediate-mass giant HD 175679.
Precise radial velocity data of the star from Xinglong Station and Okayama Astrophysical Observatory (OAO)
    revealed a Keplerian velocity variation with an orbital period of $1366.8 \pm 5.7$ days,
    a semiamplitude of $380.2 \pm 3.2 \mathrm{m\ s^{-1}}$,
    and an eccentricity of $0.378 \pm 0.008$.
Adopting a stellar mass of $2.7 \pm 0.3\ M_\odot$,
    we obtain the minimum mass of the HD 175679 b is $37.3 \pm 2.8\ M_\mathrm{J}$,
    and the semimajor axis is $3.36 \pm 0.12\ \mathrm{AU}$,
This discovery is the second brown dwarf companion candidate
    from a joint planet-search program between China and Japan.
}

   \authorrunning{L. Wang et al.}            
   \titlerunning{A Substellar Companion to the Intermediate-mass Giant HD 175679}  

   \maketitle

%
%
\section{Introduction}           
\label{sect:intro}

Brown dwarfs are widely known as ``failed stars'', with masses falling between the deuterium-burning
    limit ($\sim13\,M_\mathrm{J}$) and the hydrogen-burning limit ($\sim80\,M_\mathrm{J}$).
A brown dwarf with a mass of $15\,M_\mathrm{J}$ and a separation of $\sim3\,\mathrm{AU}$
    in a circular orbit around a solar-mass star
    causes a radial-velocity semi-amplitude of stellar motion
    of above $200\,\mathrm{m\,s^{-1}}$,
    if seen along the orbital plane,
    which is easy to be detected with precise radial velocity techniques.
However, compared with the number of planetary and stellar companions,
    the brown dwarf-mass companions revealed by Doppler surveys are rare.
Grether \& Lineweaver (2006) estimated that less than 1\% of Sun-like stars harbor brown dwarf companions.
This rate is significantly lower than that of harboring stellar ($11\pm3\%$) or giant planetary
    ($5\pm2\%$) companions (Grether \& Lineweaver, 2006).
Marcy \& Butler (2000) also reported only 0.5\% of main sequence stars
    have brown dwarf-mass companions within 3 AU. 
Such a deficit between the planetary and stellar mass in the mass distribution of companions
    is called the ``brown dwarf desert''.
The paucity of brown dwarf companions may imply two distinct formation mechanisms:
    core-accretion model (e.g. Ida \& Lin 2004, Alibert et al. 2005),
    which is thought to be the main mechanism of giant planet formation
    (e.g. Fischer \& Valenti 2005, and references therein),
    and disk instability model (e.g. Boss 1997),
    which may dominate the formation processes of brown dwarf or substellar companions.

In the past decades, 7 brown dwarf-mass companions and more than 20 planetary companions
    around intermediate-mass stars have been detected by several Doppler survey programs
    (e.g. Frink et al. 2002; Sato et al. 2003, 2008a, 2008b; Setiawan et al. 2003; Hatzes et al. 2005;
     Lovis \& Mayor 2007; Niedzielski et al. 2007; Johnson et al. 2007; Liu et al. 2008, 2009;
     Omiya et al. 2009; Han et al. 2010).
Although the number is still small compared with those around solar-mass stars,
    some remarkable properties have already been revealed,
    providing important clues on the physical properties of the protoplanetary disks.
For instance, the planet occurrence rate around intermediate-mass stars ($1.3\sim1.9\ M_\odot$)
    is 10-15\% (D{\" o}llinger et al. 2009), which is significantly higher than that
    around solar-mass stars (e.g. Cumming et al. 2008).
This can be interpreted by the higher surface densities 
    of protoplanetary disks around more massive stars (Ida \& Lin, 2005).
Nearly all the substellar companions discovered around intermediate-mass giants
    have semimajor axes larger than 0.6 AU.
The lack of inner planets can be explained by the engulfment by the central stars
   due to the tidal torque during the RGB phase (Sato et al. 2008a), 
   or formed primordially (Burkert \& Ida, 2007).
The planet-metallicity correlation for planets around solar-type stars does not seem to exist
    for those around intermediate-giants (Pasquini et al. 2007; Takeda et al. 2008),
    and therefore constrain the planet formation model (e.g. Ida \& Lin 2004, Boss 1997).

In this paper, we report the detection of a brown dwarf-mass companion candidate
    to the intermediate-mass giant HD 175679.
This is the second brown dwarf candidate
    and the third discovery of the China-Japan planet search program
    carried out at Xinglong Station (National Astronomical Observatories, China)
    and Okayama Astrophysical Observatory (OAO, Japan).


\section{Observations}
\label{sect:Obs}
\subsection{OAO observations}

The Okayama Planet Search Program started in 2001. The program has been carrying out a precise
Doppler planet survey of 300 G and K giants using the 1.88m telescope with the High Dispersion
Echelle Spectrograph (HIDES: Izumiura 1999) at OAO.
In 2007 December, HIDES was upgraded from a single CCD (2 K $\times$ 4 K) to a mosaic of three CCDs,
    which can simultaneously cover a wavelength range of 3750-7500 \AA\, using the RED cross-disperser.
We set the slit width to 200 $\mathrm{{\mu}m}$ (0.76"),
    giving a spectral resolution ($\lambda/\Delta\lambda$) of 67,000 with 3.3 pixels sampling,
    and we use an iodine absorption cell ($\mathrm{I}_2$ cell: Kambe et al. 2002) for precise wavelength calibration.
The reduction of the echelle spectra is performed using the IRAF\footnote{
    IRAF is distributed by the National Optical Astronomy Observatory,
    which is operated by the Association of Universities for Research in Astronomy, Inc.,
    under cooperative agreement with the National Science Foundation.}
    software package in the standard manner.
The $\mathrm{I}_2$-superposed (``star+$\mathrm{I}_2$'') spectra are modeled based on the algorithm
    given by Sato et al. (2002).
The stellar template used for radial velocity analysis is extracted by deconvolving an instrumental
profile, which is determined by a B-star+$\mathrm{I}_2$ spectrum, from a pure stellar spectrum taken without $\mathrm{I}_2$ cell.
The Doppler precision is less than $6\ \mathrm{m\ s^{-1}}$ over a time span of 9 yr.
We used the stellar spectra without $\mathrm{I}_2$ cell for abundance analysis
    (e.g. Takeda et al. 2008, Liu et al. 2010).

\subsection{Xinglong Observations}
To extend the Okayama Planet Search Program,
    the Xinglong Planet Search Program started in 2005 under a framework of international collaboration
    between China and Japan.
About 100 G-type giants with magnitudes of $6.0<V<6.2$ are being monitored with the 2.16 m telescope
    and the Coud\'e Echelle Spectrograph (CES: Zhao \& Li 2001) at Xinglong.
A part of the sample is also being monitored at OAO in order to confirm the stars' radial-velocity
    variability independently and HD 175679 presented in this paper is included
    in the sample observed both at Xinglong and OAO.
The iodine absorption cell attached to CES is a copy of that for HIDES at OAO.
We use the blue arm of CES for precise radial velocity measurements,
    which covers a wavelength range of 3900-7260 \AA\ with a spectral resolution of ~40,000
    by 2 pixel sampling.
Due to the small format of the CCD (1 K $\times$ 1 K, with a pixel size of 24 $\times$ 24 $\mathrm{{\mu}m}^2$),
    only about 470 \AA\ is available for precise radial velocity measurements.
The modeling of the star+$\mathrm{I}_2$ spectra and template extraction is based on the method by
    Sato et al. (2002),
    giving a radial velocity precision of $20-25\ \mathrm{m\ s^{-1}}$ over a time span of 3 yr,
    which is limited by the low resolution of the spectrograph and the narrow wavelength coverage of the CCD.
In March 2009, the CCD was replaced by Princeton Instrument's VersAarray:2048B equipped with
    an e2v CCD42-40 image sensor having a pixel size of 13 $\times$ 13 $\mathrm{{\mu}m}^2$, which was provided
    by the National Astronomical Observatory of Japan (NAOJ).
The sampling rate is increased to about 3.9 pixels,
        although the wavelength coverage nearly does not change.
Radial velocity analysis for the data with the new CCD
    is basically the same as that for the data with the old CCD,
    but we used the stellar template obtained with the OAO data for the
    analysis of the Xinglong new data. We achieved a Doppler precision of about
    $15\ \mathrm{m\ s^{-1}}$
    with a typical S/N of 150-200 over a time span of 1 yr.

\section{Stellar Properties}

HD 175679 (HR 7144, HIP 92968, BD +02 3730) is a G8III star
    with $V=6.14$, $B-V=0.96$,
    and a {\em Hipparcos} parallax of $\pi = 6.23 \pm 0.8\ \mathrm{mas}$ (ESA 1997),
    giving the distance of $161\pm21\,\mathrm{pc}$ from the Sun,
    and the absolute magnitude $M_\mathrm{V}=0.52$.
The physical parameters ($T_\mathrm{eff}$, [Fe/H], $\log{g}$, $v_\mathrm{t}$, and$M_*$)
    are taken from Liu et al. (2010), who used the stellar spectra
    obtained without $\mathrm{I}_2$ cell at OAO.
The effective temperature ($T_\mathrm{eff}=4844\pm100\ \mathrm{K}$)
    was derived from color index $B-V$ and the empirical calibration relations of Alonso et al. (1999),
    and the metallicity [Fe/H]=-0.14 was derived from the equivalent widths
    measured from the $\mathrm{I}_2$-free spectrum.
Surface gravity, $\log g=2.59\pm0.10$, was determined via {\em Hipparcos}
    parallax (ESA 1997).
The stellar mass $M_*=2.7\pm0.3\ M_\odot$ was estimated from
    the Yonsei-Yale stellar evolutionary tracks (Yi et al. 2003).
Microturbulent velocity $v_\mathrm{t}=1.4\pm0.2\ \mathrm{km\ s^{-1}}$
    was obtained by forcing Fe I lines with different strengths to give
    the same abundances.
The stellar parameters are summarized in Table \ref{tbl:stellar}.

\begin{table}
  \begin{center}
  \caption{Stellar Parameters of HD 175679}\label{tbl:stellar}
    \begin{tabular}{ll}
      \hline
      \hline
      Parameter & Value \\
      \hline
      Spectral Type         & G8III         \\
      $\pi\ (\mathrm{mas})$ & $6.23\pm0.80$ \\
      $V$                   & 6.14          \\
      $B-V$                 & 0.961         \\
      $M_V$                 & 0.52          \\
      $BC$                  & -0.318        \\
      $T_\mathrm{eff}$ (K)  & $4844\pm100$  \\
      $\log g$              & $2.59\pm0.10$ \\
      $\mathrm{[Fe/H]}$     & $-0.14\pm0.10$\\
      $v_\mathrm{t}\ (\mathrm{km\ s^{-1}})$  & $1.4\pm0.2$   \\
      $L\ (L_\odot)$        & $66\pm17$     \\
      $R\ (R_\odot)$        & $11.6\pm1.6$  \\
      $M\ (M_\odot)$        & $2.7\pm0.3$   \\
      Age\ (Gyr)            & $0.5^{+0.4}_{-0.2}$ \\
      \hline
    \end{tabular}
  \end{center}
\end{table}

\section{Radial Velocities and Orbital Solution}

\begin{figure}
   \centering
   \includegraphics[width=8cm, angle=0]{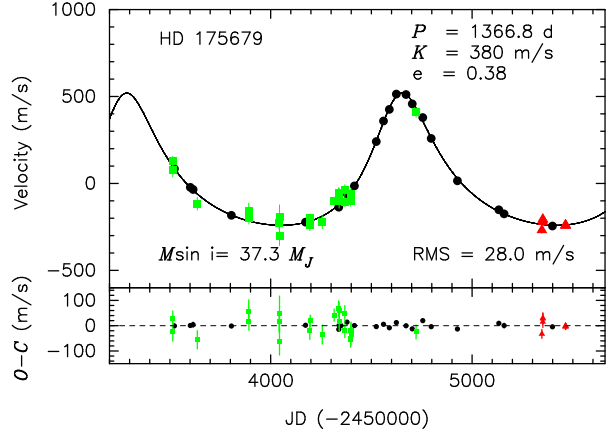}
   \caption{Radial velocities of HD 175679 observed at OAO (black circles)
            and Xinglong with old CCD (green squares)
            and new CCD (red triangles).
            The Keplerian orbit (solid line) was determined using
            both the OAO and Xinglong data.}
   \label{fig:rv-curve}
\end{figure}

The observation of HD 175679 was started at OAO and Xinglong Station in 2005.
Over a time span of 5 years, we collected a total of 22 radial velocity data points at OAO
   with a typical S/N of 250 and 31 data points at Xinglong
   (23 with old CCD and 8 with new CCD) with a typical S/N of 150-200.
The radial velocity points are shown in Figure \ref{fig:rv-curve} and listed in Table
    \ref{tbl:obs-all} together with their estimated uncertainties.
The best-fit Keplerian orbit was derived from both the OAO and Xinglong data
    using a least-squares method,
    and is shown in Figure \ref{fig:rv-curve} overplotted on the velocity data points.
We applied the velocity offset of $-$177 m s$^{-1}$ and $-$318 m s$^{-1}$
    to Xinglong old and new data respectively, relative to the OAO data in order to minimize
    reduce $\chi$-squared $(\sqrt{\chi _{\nu }^2})$ when fitting a Keplerian model to the combined
    Xinglong and OAO data.
The orbital parameters are listed in Table \ref{tbl:orbital},
    and their uncertainties were estimated using a bootstrap Monte-Carlo approach,
    subtracting the theoretical fit,
    scrambling the residuals,
    adding the theoretical fit back to the residuals and then refitting.

The radial velocity variability can be well fitted as a Keplerian orbit with
    period $P=1366.8\pm 5.7$ days,
    a velocity semiamplitude $K_1=380.2\pm 3.2$ m s$^{-1}$,
    and an eccentricity $e=0.378\pm 0.008$.
Adopting a stellar mass of $2.7\pm0.3\,M_\odot$ (Liu et al. 2010), we obtain for the companion
a minimum mass 
    $m_2\sin i=37.3\pm2.8 \,M_\mathrm{J}$ and a semimajor axis $a=3.36\pm 0.12\,\mathrm{AU}$.
Overall RMS scatter of the residuals was 28.0 m s$^{-1}$, which was due to the low precision
of the Xinglong old data. If we only use the OAO data, the RMS scatter is decreased to
8.4 m s$^{-1}$,
which is consistent with the radial velocity jitter
    ($\sim6\,\mathrm{m/s}$) due to stellar oscillations estimated
    using the scaling relations of Kjeldsen \& Bedding (1995).

\begin{longtable}{rrrc}
  \caption{Radial Velocities of HD 175679}\label{tbl:obs-all}\\
  \hline
           JD      & Radial velocity        & Error       & Observatory           \\
      ($-$2,450,000) & ($\mathrm{m\,s^{-1}}$) & ($\mathrm{m\,s^{-1}}$) & \\
  \hline
  \endfirsthead
  \caption{continued.}\\
  \hline
           JD      & Radial velocity        & Error       & Observatory           \\
      ($-$2,450,000) & ($\mathrm{m\,s^{-1}}$) & ($\mathrm{m\,s^{-1}}$) & \\
  \hline
  \endhead
  \hline
  \endfoot
3514.29867 & 76.7 & 40.2 & Xinglong (old) \\ 
3514.31969 & 126.6 & 29.8 & Xinglong (old) \\ 
3522.16608 & 84.5 & 6.4 & OAO \\ 
3601.15072 & $-$22.3 & 6.9 & OAO \\ 
3615.14723 & $-$34.8 & 10.0 & OAO \\ 
3636.04337 & $-$116.4 & 36.4 & Xinglong (old) \\ 
3805.33702 & $-$183.1 & 6.1 & OAO \\ 
3891.30602 & $-$160.4 & 47.0 & Xinglong (old) \\ 
3892.31025 & $-$198.8 & 35.4 & Xinglong (old) \\ 
4042.02550 & $-$225.7 & 40.4 & Xinglong (old) \\ 
4043.94178 & $-$191.9 & 68.6 & Xinglong (old) \\ 
4043.96374 & $-$302.3 & 54.4 & Xinglong (old) \\ 
4173.34589 & $-$222.3 & 5.0 & OAO \\ 
4194.34042 & $-$237.4 & 33.8 & Xinglong (old) \\ 
4196.36671 & $-$197.3 & 22.7 & Xinglong (old) \\ 
4256.27293 & $-$222.8 & 37.8 & Xinglong (old) \\ 
4315.14228 & $-$102.6 & 24.0 & Xinglong (old) \\ 
4337.08455 & $-$53.8 & 28.7 & Xinglong (old) \\ 
4338.03334 & $-$136.7 & 4.8 & OAO \\ 
4339.99129 & $-$117.3 & 4.5 & OAO \\ 
4340.06605 & $-$59.6 & 32.7 & Xinglong (old) \\ 
4340.07325 & $-$103.4 & 38.5 & Xinglong (old) \\ 
4348.98057 & $-$109.7 & 4.5 & OAO \\ 
4368.01697 & $-$39.1 & 31.8 & Xinglong (old) \\ 
4368.03145 & $-$105.5 & 29.0 & Xinglong (old) \\ 
4379.03102 & $-$58.2 & 5.5 & OAO \\ 
4396.96580 & $-$95.7 & 35.4 & Xinglong (old) \\ 
4397.94288 & $-$99.2 & 22.8 & Xinglong (old) \\ 
4397.95442 & $-$62.7 & 27.2 & Xinglong (old) \\ 
4398.97764 & $-$76.1 & 28.2 & Xinglong (old) \\ 
4398.99416 & $-$62.7 & 34.6 & Xinglong (old) \\ 
4415.89851 & $-$14.3 & 5.0 & OAO \\ 
4524.34441 & 241.0 & 5.7 & OAO \\ 
4560.32605 & 359.0 & 4.8 & OAO \\ 
4589.27815 & 426.0 & 4.7 & OAO \\ 
4624.11627 & 514.0 & 5.7 & OAO \\ 
4672.08312 & 511.3 & 5.1 & OAO \\ 
4703.09453 & 457.6 & 4.4 & OAO \\ 
4722.06738 & 408.2 & 27.8 & Xinglong (old) \\ 
4754.97696 & 379.1 & 4.8 & OAO \\ 
4796.88244 & 258.7 & 4.9 & OAO \\ 
4927.32075 & 15.1 & 5.0 & OAO \\ 
5132.92161 & $-$151.7 & 4.6 & OAO \\ 
5158.93295 & $-$174.9 & 9.2 & OAO \\ 
5346.30620 & $-$267.9 & 17.9 & Xinglong (new) \\ 
5351.26229 & $-$208.5 & 21.9 & Xinglong (new) \\ 
5351.28343 & $-$220.8 & 22.3 & Xinglong (new) \\ 
5351.30453 & $-$205.6 & 20.3 & Xinglong (new) \\ 
5399.05791 & $-$244.7 & 4.5 & OAO \\ 
5464.00346 & $-$242.4 & 11.5 & Xinglong (new) \\ 
5464.02484 & $-$239.0 & 10.7 & Xinglong (new) \\ 
5464.04632 & $-$241.5 & 12.0 & Xinglong (new) \\ 
5464.06775 & $-$239.4 & 10.4 & Xinglong (new) \\ 
\end{longtable}

\begin{table}
  \caption{Orbital Parameters of HD 175679 Determined from
           both Xinglong and OAO Data}
  \label{tbl:orbital}
  \begin{center}
    \begin{tabular}{ll}
      \hline
      \hline
      Parameter & Value \\
      \hline
      $P\ (\mathrm{days})$                 & $1366.8\pm$5.7   \\
      $K_1\ (\mathrm{m\ s^{-1}})$          & $ 380.2\pm$3.2   \\
      $e$                                  & $  0.378\pm$ 0.008  \\
      $\omega\ (\mathrm{deg})$             & $ 346.4\pm$1.3   \\
      $T_p\ (\mathrm{JD}-2,450,000)$       & $3263.9\pm$ 7.6  \\
      $a_1\sin i\ (10^{-3}\mathrm{AU})$    & $  44.23\pm$ 0.43 \\
      $f_1(m)\ (10^{-7}M_\odot)$           & $  61.8\pm$ 1.7  \\
      $M_p\sin i\ (M_\mathrm{J})$          & $  37.3\pm$ 2.8 \\
      $a\ (\mathrm{AU})$                   & $   3.36\pm$ 0.12  \\
      $N_\mathrm{obs}$                     &     53        \\
      $\mathrm{RMS}\ (\mathrm{m\ s^{-1}})$ &    28.0       \\
      $\mathrm{Reduced}\ \sqrt{\chi _{\nu }^2}$   &   1.5        \\
      \hline
    \end{tabular}
  \end{center}
\end{table}

\section{Line Shape Analyses}
\begin{figure}
   \centering
   \includegraphics[width=8cm, angle=-90]{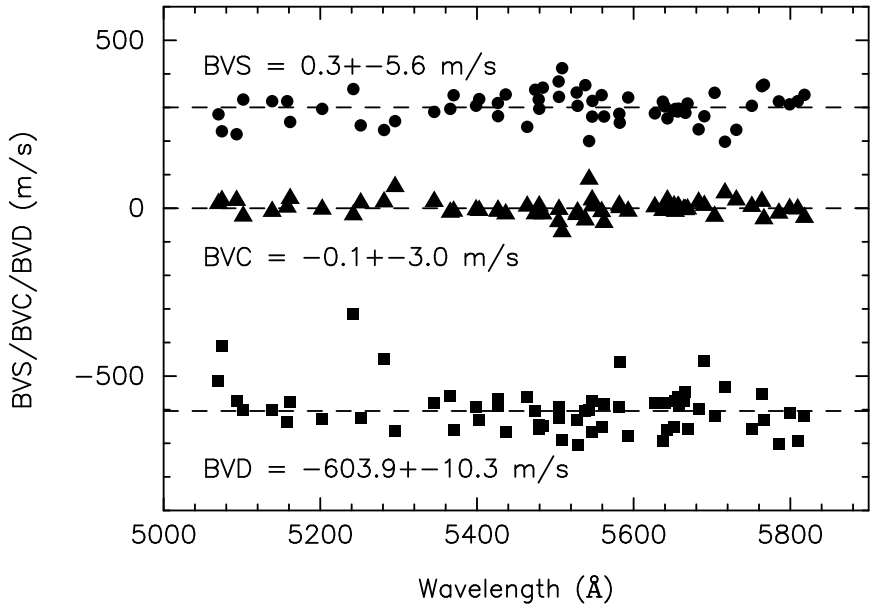}
   \caption{Bisector quantities of the cross-correlation profiles
            between the templates of HD 175679 at peak and valley phases
            of the observed radial velocities.
            The bisector velocity span (BVS, circles),
            bisector velocity curvature (BVC, triangles),
            and bisector velocity displacement (BVD, squares)
            with average values (dashed lines) and standard errors
            are shown in the figure.}
   \label{fig:bvs}
\end{figure}

We performed a spectral line-shape analysis to investigate other causes
    that could produce apparent radial-velocity variations,
    such as rotational modulation and pulsation.
The high-resolution $\mathrm{I}_2$-free stellar templates at the peak and valley
    phases of the observed radial velocities are extracted from
    several star + $\mathrm{I}_2$ spectra obtained at OAO.
Cross-correlation profiles of the templates were calculated for about 80
    spectral segments (4-5\,\AA\,width each) in which severely blended
    lines or broad lines were excluded.
We calculated three bisector quantities for the cross-correlation profile
    of each segment:
    the velocity span (BVS), which is the velocity difference between two
    flux levels of the bisecter,
    the velocity curvature (BVC), which is the difference in the velocity
    span of the upper half and lower half of the bisector,
    and the velocity displacement (BVD), which is the average of the bisector
    at three different flux levels.
We used flux levels of 25\%, 50\%, and 75\% of the cross-correlation profile
    to calculate the above three quantities,
    and the results are plotted in Figure \ref{fig:bvs}.
Both BVS and BVC for HD 175679 are identical to zero ($+0.3\,\mathrm{m\,s^{-1}}$
    and $-0.1\,\mathrm{m\,s^{-1}}$ on average, respectively),
    which means the cross-correlation profiles are symmetric.
The average BVD of $-603.9\,\mathrm{m\,s^{-1}}$ is consistent with the velocity difference between the two
    templates at the peak and valley phases of the observed radial velocities.
As a result, we concluded that the observed radial velocity variations are due to
    parallel shifts of the spectral lines caused by orbital motion.

\section{Summary and Discussion}
We report the discovery of a brown dwarf-mass companion candidate
    to the intermediate-mass giant HD 175679.
This is the second brown dwarf discovered by the joint China-Japan planet search program.
It also need to be emphasized that the unknown orbital inclination
    $i$ leaves the true mass of HD 175679 b uncertain.
If the orbit is randomly oriented, there is a 10\% chance that
   the true mass exceeds $80\,\mathrm{M_J}$ ($i<28^\circ$),
    which is the border between brown dwarf and stellar mass regimes.

\begin{table}
  \caption{Planetary and Stellar Parameters of Discovered Brown Dwarf Companions
           ($13\,M_\mathrm{J}<M_\mathrm{p}\sin{i}<80\,M_\mathrm{J}$) to Intermediate-mass
           Giants}
  \label{tbl:all-bds}
  \begin{center}
    \begin{tabular}{lcccccccc}
      \hline
      \hline
      Planet&$M_\mathrm{p}\sin{i}$& $a$   & $P$    & $e$ &$M_*$      &$M_\mathrm{p}\sin{i}/M_*$& [Fe/H]&Reference\\
            &($M_\mathrm{J}$)     &(AU)&(day)& &($M_\odot$)&($M_\mathrm{J}/M_\odot$) & &    \\
      \hline
      HD 13189 b         & 14   & 1.5-2.2 &472 & 0.27 & 2-6 & 2.3-7 &       & Hatzes et al. (2005)       \\
      NGC 4349 No. 127 b & 19.8 & 2.38    &678 & 0.19 & 3.9 & 5.1   & -0.12 & Lovis \& Mayor (2007)     \\
      11 Com b           & 19.4 & 1.29    &326 & 0.23 & 2.7 & 7.2   & -0.35 & Liu et al. (2008)         \\
      BD +20 2457 b      & 21.42& 1.45    &380 & 0.15 & 2.8 & 7.7   & -1.00 & Niedzielski et al. (2009) \\
      BD +20 2457 c $^*$ & 12.47& 2.01    &622 & 0.18 & 2.8 & 4.5   & -1.00 & Niedzielski et al. (2009) \\
      HD 119445 b        & 37.6 & 1.71    &410 & 0.08 & 3.9 & 9.6   & +0.04 & Omiya et al. (2009)       \\
      HD 180314 b        & 22   & 1.4     &396 & 0.26 & 2.6 & 8.5   & +0.20 & Sato et al. (2010)        \\
      HD 175679 b        & 37.3 & 3.4     &1367& 0.38 & 2.7 & 13.8  & -0.14 & This paper                \\
      \hline
    \end{tabular}
  \end{center}
      $^*$ BD +20 2457 c has a minimum mass $M_\mathrm{p}\sin{i} = 12.47\,M_\mathrm{J}$,
      very closed to the lower mass limit ($\sim13\,M_\mathrm{J}$) of brown dwarf.
      Here we listed BD +20 2457 c as a brown dwarf companion.
\end{table}

\begin{figure}
   \centering
   \includegraphics[width=8cm, angle=0]{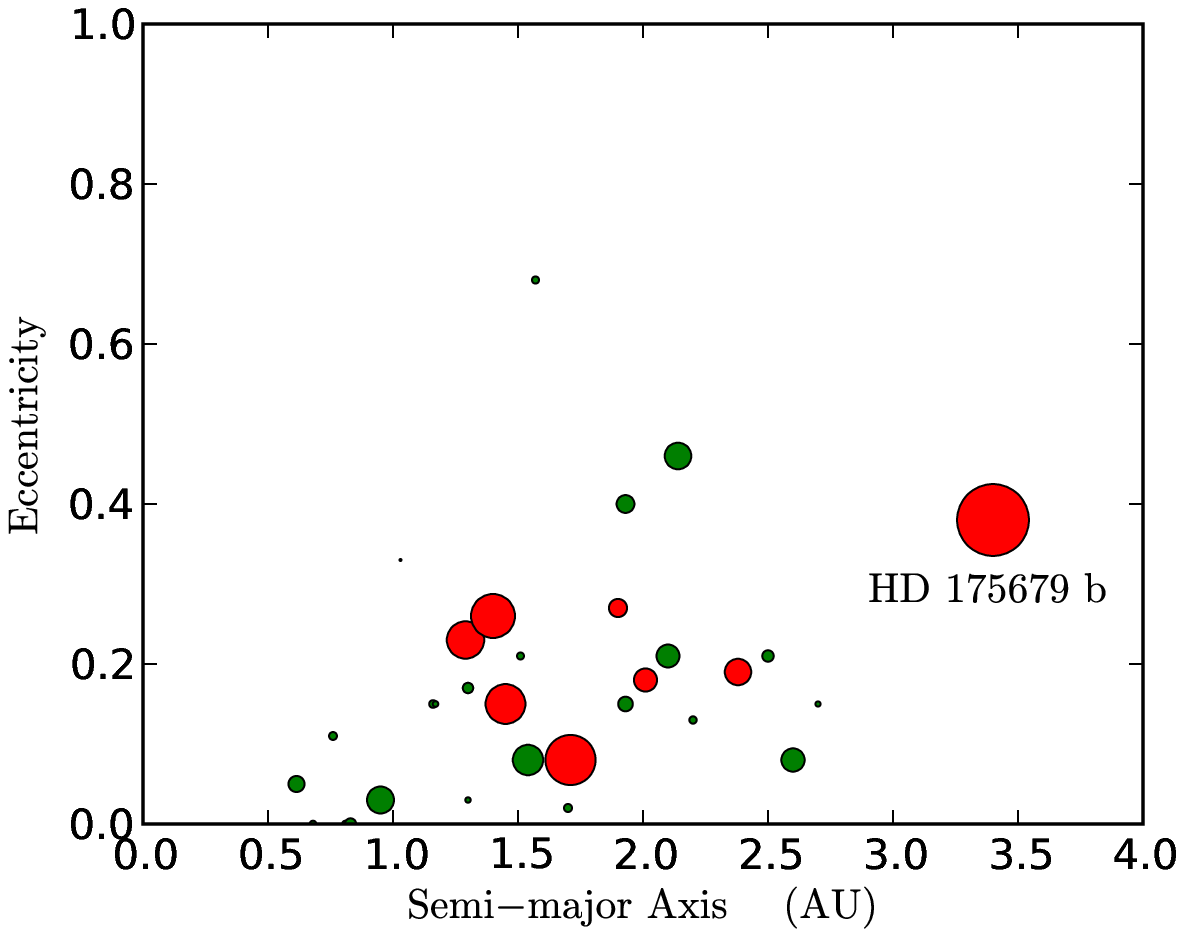}
   \caption{Semi-major axes versus eccentricity of substellar companions
            around evolved intermediate-mass stars ($1.5\ M_\odot\le M\le5\ M_\odot$).
            Green circles represent planetary companions,
            and red circles represent brown dwarf-mass companions,
            as listed in Table \ref{tbl:all-bds}.
            Radius of circles are proportional to their companion-to-host mass ratios
            ($M_\mathrm{p}\sin{i}/M_*$).
            }
   \label{fig:a-e}
\end{figure}

In Table \ref{tbl:all-bds},
    we listed parameters of discovered brown dwarf-mass companions around intermediate-mass giants
    and properties of their host stars.
The semi-major axes versus eccentricities of brown dwarf companions above are plotted in Figure \ref{fig:a-e},
    together with those of planetary companions around intermediate-mass stars ($1.5\ M_\odot\le M\le5\ M_\odot$).
The parameters of companions and host stars are taken from Table \ref{tbl:all-bds}
    and {\em The Extrasolar Planets Encyclopedia}
    \footnote{http://exoplanet.eu/, retrieved on 09/May/2010.}.
As seen in the figure, brown dwarf-mass companions reside in orbits with $a\ge 1.3$ AU, while
planetary ones exist in more inner orbits with $a\ge 0.6$ AU. This may reflect the different history
of formation and evolution between brown dwarfs and planets.

HD 175679 b has a minimum mass of $37.3\,M_\mathrm{J}$ orbiting an evolved star with $2.7\,M_\odot$,
    with a period of 1367 days.
As the 8th brown dwarf-mass companion candidate to intermediate-mass giants
    (Hatzes et al. 2005; Lovis \& Mayor 2007; Liu et al. 2008; Omiya et al. 2009;
     Niedzielski et al. 2009; Sato et al. 2010),
    HD 175679 b is somewhat unique in some respects.
Although it has been well known that more massive stars tend to host
    more massive planets or brown dwarfs
    than lower-mass stars (Lovis \& Mayor 2007; Johnson et al. 2010),
    HD 175679 b has the largest companion-to-host mass ratio
    ($M_\mathrm{p}\sin{i}/M_* = 13.8\,M_\mathrm{J}/M_\odot$)
    among those discovered brown dwarfs around intermediate-mass giants,
    and hence fall in the region (a) in Figure 5 of Omiya et al. (2009),
    which is proposed to be a paucity of brown dwarf-mass companion around 
    stars with $M_* = 1.5-2.7\,M_\odot$.

Furthermore, HD 175679 b is the first brown dwarf candidate
    with semimajor axis $a>2.5\,\mathrm{AU}$,
    and eccentricity $e>0.3$ around evolved stars,
    displaying the diversity of substellar companions falling in the `brown dwarf desert' regime.
Brown dwarfs are thought to form by gravitational collapse
    (Bonnell \& Bastien 1992; Bate 2000),
    or gravitational instability in protostellar disks
    (Boss 2000; Rice et al. 2003).
The former scenario favors a wide variety of orbital eccentricity
    and small differences in mass between the host stars and their companions,
    which is contrary to the previous discoveries
    (e.g. Omiya et al. 2009; Sato et al. 2010).
For the gravitational instability scenario,
    analytical models (e.g. Rafikov 2005, Matzner \& Levin 2005)
    and numerical simulations (e.g. Stamatellos \& Whitworth 2008)
    suggest that giant planets are formed at far more distant places
    ($\gtrsim 10\,\mathrm{AU}$),
and a high eccentricity can be excited (e.g. Muto et al. 2011) in this scenario.
On the other hand, super-massive companions with $M_\mathrm{p}>10\,M_\mathrm{J}$
    can also form by core-accretion
    scenario in protoplanetary disks  (Ida \& Lin 2004; Alibert et al. 2005).
But the wide metallicity span ($\mathrm{[Fe/H]}=-1.0\sim+0.2$) of host stars of
    discovered brown-mass companions implies they are inconsistent with the prediction
    of the core-accretion scenario that
    the probability of harboring planet is sensitive to the metallicity of a central star.
Besides, the giant planets orbiting metal-poor stars discovered by Santos et al. (2010) and
    Moutou et al. (2011) suggest that long period giant planets are not rare around
    low-metallicity stars.
It is difficult to say which scenario dominates the formation of brown dwarf mass companions
    due to the small number of discovered objects.
However, the relatively large semimajor axis and relatively high eccentricity of
    HD 175679 b make it an important supplement to the parametric distribution
    of known brown dwarf companions.
It also need to be emphasized that HD 175679 b has the longest orbital period (1367 days)
    among those ever discovered brown dwarf-mass companions to intermediate-mass giants.
Does the long period companion tend to have higher eccentricity?
Ongoing projects and future discoveries will lead to better understanding and characterizing 
    properties of substellar companions falling in the brown dwarf desert.

\begin{acknowledgements}
This research is based on data collected at Xinglong Station,
    which is operated by the National Astronomical Observatories, Chinese Academy of Sciences,
    and Okayama Astrophysical Observatory,
    which is operated by the National Astronomical Observatory of Japan (NAOJ).
We thank Dr. Xiaojun Jiang, Hongbin Li, Feng Xiao, and Junjun Jia for their expertise
    and support in the Xinglong observations.
We are grateful to all staff members of OAO for their support during the observations.
This work was funded by the National Natural Science Foundation of China
    under grants 10821061 and 10803010 and Japan Society for the Promotion of Science
    under grant 08032011-000184
    in the framework of Joint Research Project between China and Japan.
This research has made use of the SIMBAD database,
    operated at CDS, Strasbourg, France.

\end{acknowledgements}

\label{lastpage}

\end{document}